\documentclass[twocolumn,amsmath,amssymb,prl]{revtex4}
\usepackage{graphicx}
\usepackage{epsfig}
\usepackage{epsf}
\usepackage{amssymb}
\usepackage{dcolumn}
\usepackage{bm}
\usepackage{amsmath}
\usepackage{subfigure}
\usepackage{setspace}
\usepackage{ulem} 
\usepackage{color} 
\usepackage[bookmarks=true]{hyperref}
\newcommand{\mket}[1]{\vert{)1}\rangle}
\newcommand{\mbra}[1]{\langle{)1}\vert}

\newcommand{\tr}[1]{\textrm{Tr}\left[{#1}\right]}
\newcommand{\Tr}[1]{\textrm{Tr}\left[{#1}\right]}
\newcommand{\ham}{\mathcal{H}}

\newcommand{\figref}[1]{Fig. (\ref{#1})}

\graphicspath{{../../figures/FAP/}}
\def\ThisFile{\jobname}

\begin{document}
\title {Simulations of Information Transport in Spin Chains}
\author{P. Cappellaro$^1$, C. Ramanathan$^2$ and D. G. Cory$^2$}
\affiliation{$^1$ITAMP - Harvard-Smithsonian Center for Astrophysics, Cambridge, MA 02138, USA\\ $^2$Massachusetts Institute of Technology, Department of Nuclear Science and Engineering, Cambridge, MA 02139, USA\bigskip}
\begin{abstract}
Transport of quantum information in linear spin chains has been the subject of much theoretical work. Experimental studies by nuclear spin systems in solid-state by NMR (a natural implementation of such models) is complicated since the dipolar Hamiltonian is not solely comprised of nearest-neighbor XY-Heisenberg couplings. We present here a similarity transformation between the XY-Heisenberg Hamiltonian and the grade raising Hamiltonian, an interaction which is achievable with the collective control provided by radio-frequency pulses in NMR. Not only does this second Hamiltonian allows us to simulate the information transport in a spin chain, but it also provides a means to observe its signature experimentally. 
\end{abstract}
\maketitle


In many solid-state proposals for quantum computers, one essential task is the transport of information over relatively short distances. Relying on photons requires a frequent exchange of information between solid-state and light qubits that could be too costly. Linear chains of spins coupled by the XY-Heisenberg Hamiltonian could be a viable alternative \cite{Bose,tombesi}. 
Various schemes have been proposed to obtain perfect state transfer, involving engineering of the system \cite{christandl-2004-92} or of the Hamiltonian \cite{Kay,fitzsimonsSpinChain}. Usually these schemes are based on nearest-neighbor couplings only, which enable analytical solutions of the dynamics \cite{JW}. Taking into account the full couplings \cite{KayNN,paternostro} complicates the dynamics and steps toward control sequences to reduce the couplings to nearest-neighbors  have been taken \cite{altafini,sidnnetal}. A quantum simulation of this transport mechanism could  connect these theoretical studies to  the phenomenon of spin diffusion \cite{Bloembergen} observed in dipolarly coupled spin systems. While optical lattices have been proposed as versatile quantum simulators for a variety of spin Hamiltonians \cite{atomicToolbox,cirac}, here we propose to use a system closer to the simulated one, taking advantage of the control techniques developed by the NMR community.

In this letter we show how well-known NMR pulse sequences can be used to experimentally study the transport of quantum information, in collectively controlled, room temperature linear spin chains. Quasi-1D spin systems are available in some materials like apatites (either  hydroxyapatites or fluorine containing apatites \cite{FAP2exp,FAP1exp,FAPYesi2}). We first introduce a Hamiltonian (the so-called double-quantum (DQ) Hamiltonian) that is connected to the XY-Hamiltonian by a similarity transformation and that is obtained experimentally by sequences of pulses and delays in dipolar spin systems. We show how this Hamiltonian can simulate the transport dynamics for relevant initial states, by comparing the probability of transfer for the two Hamiltonians. We show furthermore that the DQ-Hamiltonian enables the detection of successful transport, even when it is not possible to measure single nuclear spins. We conclude by discussing some of the questions that the experimental studies enabled by this method could attack.

Transport of Zeeman and dipolar energy in spin systems, caused by energy conserving spin flips, has been a long term interest in spin physics \cite{Bloembergen,waughdiffusion} and more recently in quantum information science \cite{corydiffusion1,boutisetal,greenbaum,greenbaumFF}. 
The evolution of the system induced by the secular dipolar Hamiltonian
\begin{equation}
	\label{DipHam}
 	\ham_{dip}=\sum_{ij}d_{ij}[\sigma_z^i\sigma_z^j-\frac{1}{2}(\sigma_x^i\sigma_x^j + \sigma_y^i\sigma_y^j)]	,
\end{equation}
is coherent (as proved by the observation of polarization echoes \cite{TimeReversal,MEetal,Pastawski,Pastawski06,JoonFid}), but the complexity of the interaction encodes information about the  many-body states created in observables that are not directly measurable in NMR experiments. The free-evolution transport presents the  signature of an apparent diffusive behavior, with an effective decay time $T_2^*$ much shorter than spin-lattice relaxation times.  Refocusing experiments can extend the transverse relaxation $T_2$ to much larger values, although still quite far from the theoretical limit of the longitudinal relaxation $T_1$. 
If no special assumptions are made on the values of the dipolar couplings \cite{KayNN}, spin diffusion (that is, evolution under the secular dipolar Hamiltonian)  cannot transfer the polarization from the first to the last spin with any appreciable efficiency \cite{WaughErgodic,ErnstErgodic}, even for a small number of spins in linear chains. 
On the other hand, a simpler Hamiltonian, the XY interaction, 
\begin{equation}
	\label{FFHam}
	\ham_{\textsf{xy}}=\sum_{ij}\frac{d_{ij}}{2}(\sigma_x^i\sigma_x^j + \sigma_y^i\sigma_y^j)=\sum_{ij}d_{ij}(\sigma_+^i\sigma_-^j + \sigma_-^i\sigma_+^j),
\end{equation}
has been studied extensively for perfect transport purposes, but it is not found naturally in spin systems. It has been shown that perfect transport of a state is possible only for chains up to 3 spins \cite{christandl-2004-92}, but if one can engineer the coupling strength \cite{christandlPRAetal} or add external manipulation of all the spins \cite{khanejaTransport} or even just of the spins at the chain end \cite{Giovannetti}, perfect transport is achievable for chains of arbitrary length.

Unfortunately, given the dipolar Hamiltonian \eqref{DipHam} it is impossible with only collective control over the spins to break its symmetry and obtain the XY-Hamiltonian (by eliminating the term $\propto \sigma_z^i\sigma_z^j$). This collective control, by means of a sequence of radio-frequency (rf) pulses and delays \cite{MQC}, allows us to create the DQ-Hamiltonian:
\begin{equation}
	\label{DQHam}
 \ham_\textsf{dq}=\sum_{ij}\frac{d_{ij}}{2}(\sigma_x^i\sigma_x^j - \sigma_y^i\sigma_y^j)= \sum_{i,j}d_{i,j}(\sigma^+_i\sigma^+_j + \sigma^-_i\sigma^-_j)
\end{equation}
A simple unitary transformation links the DQ- and XY-Hamiltonian if we restrict the couplings to nearest-neighbor spins only \cite{Feldman}. The transformation however cannot be obtained by an rf pulse sequence, since it is not a collective operation:
\begin{equation}
	\label{SimTransf}\begin{array}{lr}
	\ham_{\textsf{xy}}=U^\textsf{xy}_\textsf{dq}\ham_\textsf{dq}(U^\textsf{xy}_\textsf{dq})^\dag,&U^\textsf{xy}_\textsf{dq}=\exp(-i\sum_k'\sigma_x^k)
\end{array}\end{equation}
where the sum is restricted to either even or odd spins in the chain.

Since the two Hamiltonians are related by a similarity transformation, one can use the DQ-Hamiltonian (experimentally available) to simulate the dynamics of the transport Hamiltonian, provided one can prepare initial states and detect observables which are  invariant under the transformation. With these assumptions, even if the similarity transformation $U^\textsf{xy}_\textsf{dq}$ is not explicitly applied, the evolution of these particular initial states is the same. 
For thermally polarized spin systems at $T>100mK$, the most common initial state is to have just one spin polarized, $\rho(0)=\sigma_z^a$. Since this state is invariant under the  transformation $U_\textsf{dq}^\textsf{xy}$, the DQ-Hamiltonian opens the possibility to experimentally study spin transport in dipolarly coupled systems.  
Incidentally, the pure state usually considered in theoretical studies, that is, the one-spin excitation state, is not invariant under the transformation but since here we are considering an experimental implementation, mixed states are more relevant.
 
Notice that the isomorphism between the XY- and DQ-Hamiltonian is valid also in two and three dimensions (and in general on hypercubes of any dimensions, which have already been shown to allow perfect state transport under the XY interaction \cite{christandlPRAetal}). These spin systems can be thought as tree-like graphs \cite{pyramidprl}, where the first spin (a spin at one edge of the hypercube) is at the top of the tree, and has $D$ neighbors in the following level, where $D$ is the dimension of the hypercube; in general each node has $D$ links to the upper level and $D$ links to the lower level. The similarity transformation is performed by flipping all the spins of either the odd or even levels.

We study the polarization transfer  given by the XY- and DQ-Hamiltonian, when the transformation  $U^\textsf{xy}_\textsf{dq}$ is not explicitly applied (as it will be experimentally).  The dynamics is solved in terms of fermion operators and of their Fourier transform \cite{Lacelle,Feldman,SelEndChain}:
\begin{equation}
	\label{fermionicsimple}
	\begin{array}{ll}
\displaystyle	c_j=-\prod_{k=1}^{j-1}\left(\sigma^k_z\right) \sigma_j^-; &
\displaystyle	 a_k=\sqrt{\frac{2}{N+1}}\sum^{N}_{j=1}\sin{(kj)} c_j
	\end{array}
\end{equation}
where $k=\frac{\pi n}{N+1}$, $n\in$ integers.
The nearest-neighbor coupling XY-Hamiltonian is diagonal in terms of fermion operators:
\begin{equation}
	\label{FFHamFer}
	\ham_{\textsf{xy}}=d\sum_j(c^\dag_jc_{j+1}+c_jc^\dag_{j+1})=2d\sum_k\cos{(k)}\,a^\dag_ka_k
\end{equation}
In order to diagonalize the DQ-Hamiltonian, we need a further transformation to Bogoliubov operators \cite{bogoliubov}:
\begin{equation}
\label{Bogsimple}
a_k=\frac{1}{\sqrt{2}}(\gamma_k d_k +  d^\dag_{-k})
\end{equation}
where $\gamma_k\equiv\textrm{sgn}(k)$.
We finally obtain a diagonal form, 
\begin{equation}
\label{DQHamBog}
\ham_\textsf{dq}=-2d\sum_k \cos k(d^\dag_kd_k+d^\dag_{-k}d_{-k}-1),
\end{equation}
with the same eigenvalues as  the XY-Hamiltonian. 

\begin{figure}[htb]
	\centering
		\includegraphics[scale=0.38]{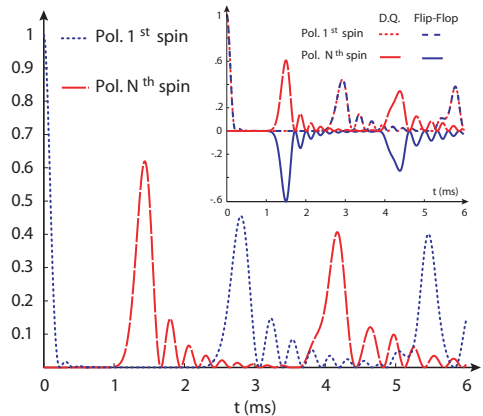}
	\caption{ \textbf{Polarization transfer} from spin one to spin N for a chain of 21 spins with nearest-neighbor coupling only; the results of evolution under double-quantum and XY-Hamiltonian are superimposed. The dipolar coupling strength considered are those found in a Fluorapatite crystal \cite{SelEndChain}. In the inset, different behavior for even or odd number chain lengths}
	\label{transport}
\end{figure}
Assuming that the polarization resides initially just on spin $a$, the initial state is: 
\begin{equation}
	\label{rho0a}
	\rho_a(0)=\frac{\openone}{2}-\frac{2}{N+1}\sum_{k,h} \sin{(ka)}\sin{(ha)}\,a^\dag_k a_h
\end{equation}
which evolves under the XY-Hamiltonian \eqref{FFHamFer} as:
\begin{equation}
	\label{rhoTaFF}
	\begin{array}{ll}
	\rho^{\textsf{xy}}_a(t)=\frac{\openone}{2}-\frac{2}{N+1}\sum_{k,h}& \sin{(ka)}\sin{(ha)} \\ & \times e^{-2id(\cos{k}-\cos{h})t}a^\dag_k a_h
	\end{array}
\end{equation}
At a time $t$, the polarization transferred to another spin $b$ is given by $\tr{\rho^{\textsf{xy}}_a(t)\sigma_z^b}$, that is:
\begin{equation}
	\label{polbff}
	P^{\textsf{xy}}_{ab}(t)=\frac{4}{(N+1)^2}\Big|\sum_k\sin{(ka)}\sin{(kb)}e^{-i\psi_k(t)}\Big|^2,
\end{equation}
where $\psi_k(t)=2\,d\,t\cos{k}$.
Evolution under the DQ-Hamiltonian with the same initial state yields: 
\begin{equation}
	\label{rhoatdq}
	\begin{array}{l}
	\rho^{\textsf{dq}}_a(t)=\frac{\openone}{2} -\frac{2}{N+1}\sum_{k,h}\sin{(ka)}\sin{(ha)}\\
\left[i\sin{\psi_k(t)} \cos{\psi_h(t)}(a^\dag_ha^\dag_{-k}-a_{-k}a_h)\right.\\
\left.+(\cos{\psi_k(t)}\cos{\psi_h(t)}\,a^\dag_ka_h+\sin{\psi_k(t)}\sin{\psi_h(t)}\,a_ha^\dag_k)\right]\end{array}
\end{equation}
 and the polarization transferred to the $b$ spin at time $t$  is
\begin{equation}
	\label{polbdq}	P^{\textsf{dq}}_{ab}(t)=\frac{4}{(n+1)^2}\textrm{Re}\left[\left(\sum_k\sin{(ka)}\sin{(kb)}e^{-i\psi_k(t)}\right)^2\right],
\end{equation}
 where $\textrm{Re}[.]$ is the real part of a complex number.
  Notice that $P^{\textsf{xy}}_{ab}(t)= \|P^{\textsf{dq}}_{ab}(t)\|$: The polarization transfer is the same if the difference $b-a$ is even, while there is a difference for $b-a$ odd [see \figref{transport}]. In the latter case the observable is no longer invariant under the similarity transformation; to recover the result $P^{\textsf{xy}}_{ab}(t)= P^{\textsf{dq}}_{ab}(t)$ one should have applied the transformation $U^\textsf{xy}_\textsf{dq}$ explicitly. 
  
More generally, if we measure a different observable than $\sigma_z^b$, we expect different behaviors whether the evolution is driven by $\ham_\textsf{dq}$ or $\ham_\textsf{xy}$. In particular the XY-Hamiltonian conserves the total magnetic number (the system remains in the zero-quantum coherence manifold), while the DQ-Hamiltonian can create multiple quantum coherence (MQC, states that show a coherence between different z-magnetic moment states). In the nearest-neighbor coupling limit, the DQ-Hamiltonian creates only zero and double-quantum coherences \cite{Feldman} and its dynamics can be solved exactly. (For couplings among all the spins $\ham_\textsf{dq}$ creates all the even coherences, as observed in MQC experiments in NMR \cite{MQC}). 
\begin{figure}[htb]
	\centering
		\includegraphics[scale=0.38]{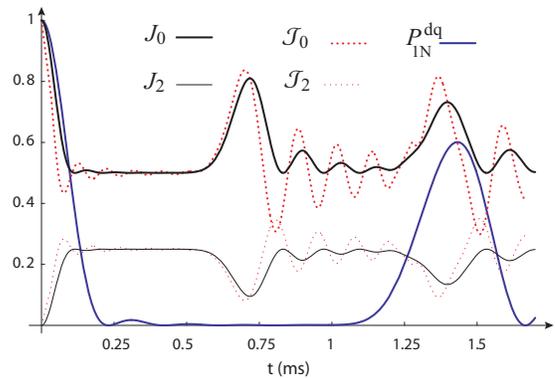}
	\caption{MQC intensities and polarization transfer in a 21-spin chain. The initial state was the one we can prepare experimentally, $\rho_0=\sigma_z^1+\sigma_z^N$ \cite{SelEndChain}.}
	\label{MQCFig}
\end{figure}

 This property of the DQ-Hamiltonian provides a means to detect the occurred transfer of polarization. The transport of polarization cannot be detected directly (unless one could introduce very strong magnetic field gradients or if one could perform single-spin detection). In the case of double-quantum dynamics, however, there is a correlation between the coherence intensities and the transport of the spin state, which arise from  boundary effects in  finite length spin chains \cite{FeldmanFiniteSize,FeldmanErnst}. 

MQC intensities can be detected in NMR, by encoding the coherence order into a phase associated with rotations around the z axis \cite{MQC, JoonPRB}. In MQC experiments, states with higher coherence orders are first prepared, for example using $\ham_\textsf{dq}$,  $\rho(0)\stackrel{\ham_{\textsf{dq}}}{\rightarrow}\rho=\sum_q \rho^{(q)}$, where q indicates the coherence order; after possible manipulations that depend on the goal of the actual experiment, the system has to be  brought back to a single-coherence state in order to be measured. The experiment is repeated $M_q$ times, each time imposing a $\varphi_k=2\pi k/M_q$ rotation around the z axis, so that different coherences acquire a phase proportional to their coherence order $\varphi_q=q\varphi$. The signal is then Fourier-transformed with respect to this phase, to obtain the intensity of each coherence. Restricting the interaction to nearest-neighbor only in 1D, the zero- and double-quantum intensities $J_\alpha(t)=\Tr{(\rho^{(\alpha)})^2}$ can be calculated analytically \cite{SelEndChain}, yielding for an initial state with only spin $a$ polarized : 
\begin{subequations}
	\label{MQCspina}
	\begin{gather}	\begin{array}{l}	J_0^a(t)=\frac{4}{(N+1)^2}\sum_{k,h}\sin{(ka)}^2\sin{(ah)}^2\\\times\cos{[\psi_k(t)+\psi_h(t)]}^2
	\end{array}
	\\
	\begin{array}{l}
	J_2^a(t) =\frac{2}{(N+1)^2}\sum_{k,h} \sin{(ka)}^2\sin{(ha)}^2\\\times \sin{[\psi_k(t)+\psi_h(t)]}^2
		\end{array}
\end{gather}
\end{subequations}

These intensities present a beating every time the polarization reaches spin $N+1-a$ and back to $a$. These  beatings  are particularly clear for the transfer from spin $1$ to spin $N$, although they would exist for the magnetization starting at any spin in the chain. If one is therefore interested in the transfer of polarization from one end of the chain to the other, it is possible to follow this transfer driven by the DQ-Hamiltonian by measuring the MQC intensities. If the measurement only allows to detect the total magnetization, the MQC intensities are then calculated from $\tr{\rho(t)^{(n)} (U_{\textsf{dq}}\sum \sigma_z U_{\textsf{dq}}^\dag)^{(n)}}$, resulting in 
\begin{subequations}
	\label{MQCspina2}
	\begin{gather}
\mathcal{J}_0^a(t)=\frac{2}{(N+1)}\sum_{k}\sin^2{(ka)}\cos^2{2\psi_k(t)}
	\\
	\mathcal{J}_2^a(t)=\frac{1}{(N+1)}\sum_{k}\sin^2{(ka)}\sin^2{2\psi_k(t)}
\end{gather}
\end{subequations}
and a less visible signature (as shown in fig. \ref{MQCFig}).

It has been shown \cite{SelEndChain} that with collective coherent and incoherent control it is possible to create the initial state $\sigma_z^1+\sigma_z^N$  (even if it is not possible to break the symmetry between the two spins at the end of the chain). In \figref{MQCFig} we show the transport and detection method results for this particular initial state, realizable experimentally. The beatings of the MQC intensities is now faster than the transfer of polarization, since the coherences start spreading out from the two opposite ends of the chain and  an extremum in the MQC intensity is created when the two waves meet at the center of the chain as well as when they bounce off the boundaries. This classical interference  occurs with a positive or negative phase, depending on the number (odd or even respectively) of spins in the chain. Every two beatings, however, the maximum of the zero-quantum intensity correspond to the transport of polarization from one end to the other one. This is the experimentally measurable signature that transport of polarization has occurred.

An experimental study of this scheme can be implemented in single crystal of fluorapatite \cite{FAP2exp,FAP1exp,FAPYesi2,SelEndChain}. These crystals present a quasi-1D structure, where fluorine spins-1/2 are arranged in linear chains, with the cross-chain couplings much smaller than the in-chain couplings (40:1 in a particular crystal orientation). Good control, long decoherence times and the availability of the desired initial state make this system very promising. In addition, a real system will provide more insight into the limitations of the nearest-neighbor coupling approximation. There already exist schemes for reducing a long-distance interaction to nearest-neighbors only \cite{altafini,sidnnetal}, but they are valid only for a restricted number of spins. On the other hand it is interesting to study the role of the long-distance couplings in accelerating or impeding the transport, as well as the cross-chain couplings. Next-nearest neighbor couplings and cross-chain couplings offer additional pathways that can result in an acceleration of information transport, which has no classical counterpart, as already observed in the transport of dipolar energy in spin diffusion \cite{greenbaum,boutisetal}. 
It will be interesting to investigate the differences between the predicted rate of transport and the experimental one, to observe the effects of the real couplings on the spin dynamics, as we move from a solvable system to systems that display fully quantum-mechanical dynamics.
 
In conclusion, we have shown how the DQ-Hamiltonian simulates the transport of polarization and enables its detection. This is made possible by the observation that not only is this Hamiltonian related to the XY interaction via a similarity transformation, but it also creates coherent states, whose intensities are correlated to the transport of polarization and can be measured experimentally. With this scheme it will be possible to study experimentally, in solid-state NMR systems, the transport of polarization beyond exactly solvable models and explore the appearance of quantum  coherence and interference effects. We anticipate that the control and measurement techniques  described in this letter will allow the simulation of more general problems in condensed matter physics. 

\textbf{Acknowledgments}.  
This work was supported in part by the National Security Agency (NSA) under Army Research Office (ARO) contracts DAAD190310125 and W911NF-05-1-0459, by DARPA and by the National Science Foundation under award 0403809 and  through a grant for the Institute of Theoretical, Atomic, Molecular and Optical Physics.

\bibliography{../UpdateBib}
\end{document}